\documentclass[a4paper]{article}

\usepackage[utf8]{inputenc}
\usepackage[T1]{fontenc}
\usepackage[english]{babel}
\usepackage[bookmarks=false,hidelinks]{hyperref} 
\usepackage{lmodern}
\usepackage[
  backend=biber,
  url=false,
  style=trad-plain,
  maxnames=4,
  minnames=3,
  maxbibnames=99,
  giveninits,
  uniquename=init]{biblatex}
\usepackage{microtype}
\usepackage{graphicx}
\usepackage{url}
\usepackage{authblk}
\usepackage{csquotes}
\usepackage{import}
\usepackage{layout}
\usepackage{color}
\usepackage{amssymb}
\usepackage{amsmath}
\usepackage{tikz}
\usepackage{tikz-uml}
\usepackage{comment}

\addto\extrasenglish{

}

\title{
\def\svgwidth{0.23\textwidth}
\begingroup%
  \makeatletter%
  \providecommand\color[2][]{%
    \errmessage{(Inkscape) Color is used for the text in Inkscape, but the package 'color.sty' is not loaded}%
    \renewcommand\color[2][]{}%
  }%
  \providecommand\transparent[1]{%
    \errmessage{(Inkscape) Transparency is used (non-zero) for the text in Inkscape, but the package 'transparent.sty' is not loaded}%
    \renewcommand\transparent[1]{}%
  }%
  \providecommand\rotatebox[2]{#2}%
  \newcommand*\fsize{\dimexpr\f@size pt\relax}%
  \newcommand*\lineheight[1]{\fontsize{\fsize}{#1\fsize}\selectfont}%
  \ifx\svgwidth\undefined%
    \setlength{\unitlength}{300.00204011bp}%
    \ifx\svgscale\undefined%
      \relax%
    \else%
      \setlength{\unitlength}{\unitlength * \real{\svgscale}}%
    \fi%
  \else%
    \setlength{\unitlength}{\svgwidth}%
  \fi%
  \global\let\svgwidth\undefined%
  \global\let\svgscale\undefined%
  \makeatother%
  \begin{picture}(1,0.99999351)%
    \lineheight{1}%
    \setlength\tabcolsep{0pt}%
    \put(0,0){\includegraphics[width=\unitlength,page=1]{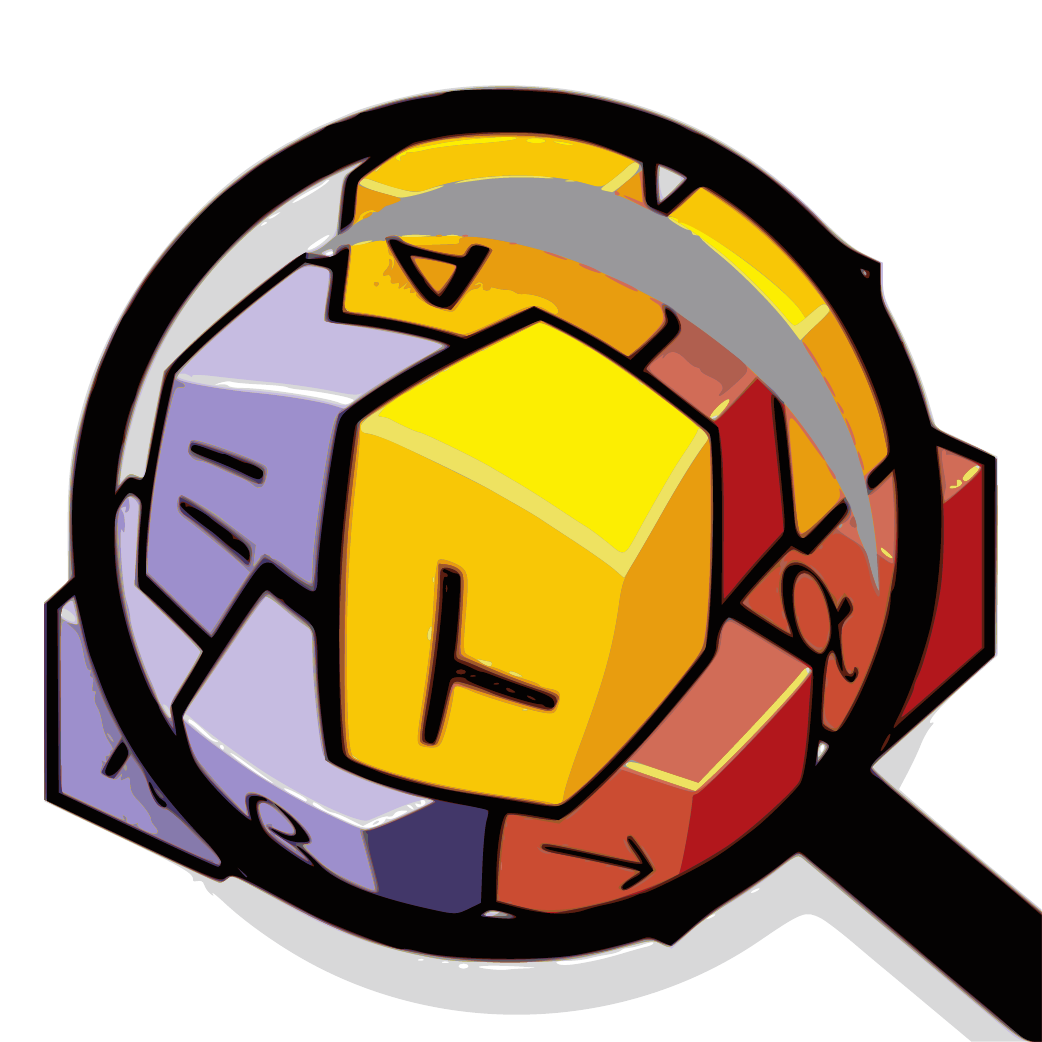}}%
  \end{picture}%
\endgroup%

\\[0.8cm]
FindFacts: A Scalable Theorem Search}
\author[1,2]{Fabian Huch}
\author[2]{Alexander Krauss}
\affil[1]{Technische Universität München}
\affil[2]{QAware GmbH}
\date{May 2020}

\bibliography{references}

\begin{document}
\maketitle



\begin{abstract}
The Isabelle Archive of Formal Proofs (AFP) has grown to over 500 articles in late 2019.
Meanwhile, finding formalizations in it has not exactly become easier.

At the time of writing,
the site-specific AFP google search\footnote{\url{https://www.isa-afp.org/search.html}}
and the Isabelle \texttt{find\_theories} and \texttt{find\_consts} commands
(that only work on imported theories)
are still the only tools readily available to find formalizations in Isabelle.

We present \emph{FindFacts}\footnote{\url{https://search.isabelle.in.tum.de}}, a novel domain-specific search tool for formal Isabelle theory content.
Instead of utilizing term unification, we solve the problem with a classical drill-down search engine.
We put special emphasis on scalability of the search system, so that the whole AFP can be searched interactively.

\end{abstract}
\section{Introduction}\label{sec:introduction}

The AFP has grown to a magnificent store of knowledge,
incorporating more than 500 articles from over 340 different authors,
which, in total, consist of nearly 2.5 million lines of Isabelle code \cite{isa-afp2020statistics}.
Due to the sheer size of the archive, finding specific formalizations becomes an increasingly challenging task.

The current solution to this problem is a site-specific Google search,
in which the theory files of Isabelle and the AFP are indexed.
However, this generic search does not exploit much of the potential that lies in searching in formal theories,
since it does not utilize the semantic content at all.
Moreover, there are multiple practical issues:
Since every release is indexed, results contain lots of near-duplicates --
a problem that only gets worse as more revisions are published.
Moreover, the order of the returned results is not very meaningful,
as the ranking is based on links on and between the theory files.

The built-in Isabelle query tools on the other hand -- \texttt{find\_theorems} and \texttt{find\_consts} --
operate directly on loaded theory content,
but can only find results of the currently active session.
Thus, they are impractical for searching the whole AFP, which is far too big to be loaded into memory.
Also, they require the user to have a good idea of the structure of entities they are looking for --
which is often not the case when looking for unknown formalizations.
We thus state the issue as follows:

\textbf{Problem.}
Finding unknown formalizations in the AFP (or large sessions of Isabelle) is hard to do within the system itself,
and has no proper tool support outside of it.

\textbf{Solution.}
To alleviate this problem,
we propose a domain-specific search system that indexes theory code as well as semantic content,
and allows drilling down on large result sets by filtering for theory-specific properties.
Our solution focuses on scalability to large data sets,
such that the whole AFP (and potentially much more) can be searched with sub-second response times.
As a trade-off, we accept to give up some of the more advanced search capabilities of \texttt{find\_theorems},
such as matching and unification.
This allows us to rely on widely-used open-source components for the actual indexing and search functionality.
Our implementation relies on Apache Solr \cite{solr82} for this purpose.
While Solr is primarily text-based and does not understand lambda calculus or term structure,
we claim that search results are helpful for many use cases.

\textbf{Contribution.}
In this paper, we lay out the architecture of our novel FindFacts tool,
and explain in detail how its components function,
as well as how the application core can be re-used and integrated into other tools,
for instance as a standalone Isabelle tool.

\textbf{Organization.}
\autoref{sec:related} covers related tools.
The FindFacts system is explained in \autoref{sec:search};
in \autoref{sec:future}, we will pragmatically break down the backlog of future changes and potential additions.

\section{Related Work}\label{sec:related}
The built-in commands \texttt{find\_theorems} and \texttt{find\_consts} find formal content in the currently loaded context.
Both support filtering by names, allowing wildcards.
\texttt{find\_theorems} can search for terms matching a term pattern,
and for theorems that act as specific deduction rules on the current goal,
for instance elimination rules.
Similarly, \texttt{find\_consts} can search for a type pattern \cite{wenzel2019isabelle}.
These tools bring up useful results when they are available in the current context,
but do not help much when the relevant material lives outside the current scope, such as a separate AFP entry.

Coq provides a command \texttt{Search} that is comparable to \texttt{find\_theorems} and has similar limitations \cite{coq2019search}.

To allow searching for content that is not currently loaded,
the verification group at NICTA contributed a simple tool called \emph{wwwfind} to Isabelle 2009-1,
where users could run queries via a web interface,
which were then executed by \texttt{find\_theorems} in an Isabelle instance with a pre-loaded image running on the server.
This approach definitely helped in some cases, but it was still limited to a specific logic image
and could not scale to the whole AFP.
The tool was removed again in Isabelle 2014 after being unmaintained for a while.

Another theorem search approach is the \emph{MathWebSearch} (MWS) system.
Its main purpose is to find results in mathematical documents;
to do that, it indexes formulas (in the \emph{MathML} format \cite{ion1998mathematical}) as well as text,
and -- in addition to search by keywords -- it allows searching for terms by unification, even for large indexes \cite{hambasan2014mathwebsearch}.
However, it is not directly usable for Isabelle theories.

\section{The FindFacts System}\label{sec:search}
The FindFacts search is an external tool with a web interface rather than an Isabelle command.
There are multiple reasons for this:
To begin, it does not have a concept of a `currently loaded context';
rather, it operates on a pre-built index all at once,
making it possible to search in a far greater scope than tools such as \texttt{find\_theorems} can.
Building this index is computationally very expensive
and thus has a very different life-cycle than an Isabelle session.
Another reason is decoupling: 
The search does not share much functionality with Isabelle.
Reducing coupling to the fast-changing Isabelle codebase greatly decreases maintenance effort,
and also minimizes external dependencies of the Isabelle system.

To be able to index the whole AFP and also complete searches instantaneously,
the search system is separated into a data indexing batch process
and a persistent data querying/retrieving component.

\subsection{Import Process}
\begin{figure}[ht]
    \centering
    \def\svgwidth{\columnwidth}
    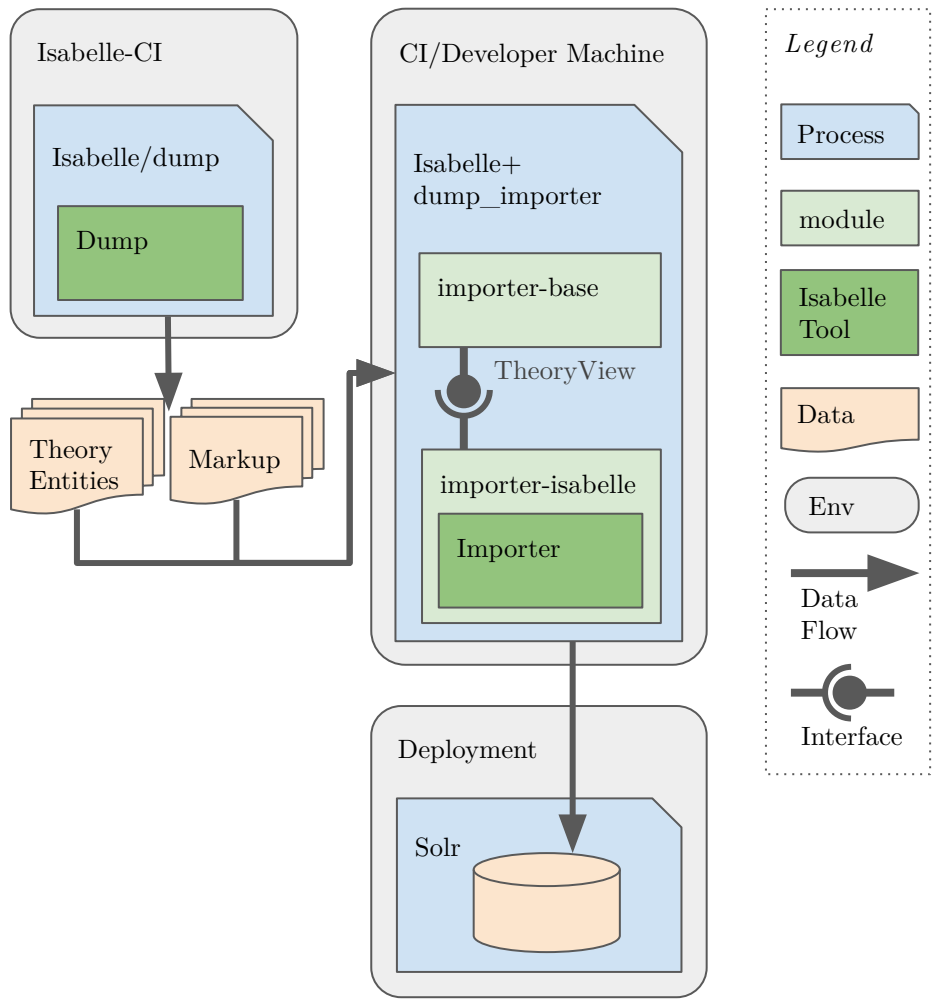
    \caption{Import architecture overview}
    \label{fig:import_architecture_overview}
\end{figure}

The import process is depicted in \autoref{fig:import_architecture_overview}.
For the resource-intensive export of semantic theory content,
the existing Isabelle \texttt{dump} tool is used.
It handles partitioning the theorem graph of the AFP into processible chunks,
dynamically loading and unloading theories
to minimize the re-building of base sessions while limiting memory consumption.
It yields theory entities as well as the semantic markup of theory sources;
further processing of these artifacts is far less demanding.

We import the resulting artifacts in a second step;
the dump provides an appropriate interface between the export and the indexing re-import.
Our external Isabelle component \texttt{dump\_importer}
(with a corresponding command-line wrapper)
is responsible for the second step.
Isabelle-specific parts of the implementation reside in the \texttt{importer-isabelle} module;
in an adapter pattern,
that module implements the \texttt{TheoryView} interface to expose theory elements.
The implementation is provided to the \texttt{importer-base} module,
which is responsible for processing of the entities and creation of the search index.
The final data model is explained in \autoref{subsec:datamodel}.
We use Apache Solr \cite{solr82} as a search database.
It is a scalable high performance in-memory document store
that is based on the Apache Lucene \cite{lucenecore} full-text search engine.
We provide an abstraction over the search server (via \texttt{SolrRepository} interface),
so that it can interchangeably be run embedded,
as a separate (remote) database,
or in a distributed cloud setup.

\subsection{Search Application}
\begin{figure}[ht]
    \centering
    \def\svgwidth{\columnwidth}
    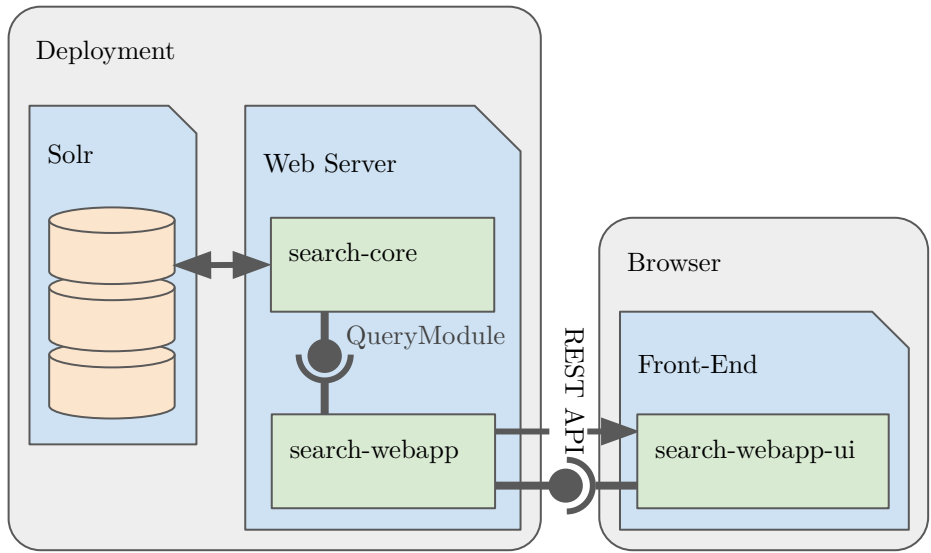
    \caption{Search application architecture; legend as in \autoref{fig:import_architecture_overview}}
    \label{fig:search_architecture_overview}
\end{figure}

Once an index has been built,
the \texttt{search-core} module can be used standalone to execute queries via its Scala interface.
The query specification is explained in \autoref{subsec:query}.
As summarized in \autoref{fig:search_architecture_overview},
our web application located in the \texttt{search-webapp} module provides a REST-based API for queries
and serves the front-end (in \texttt{search-webapp-ui}).
The back-end is built with the Play framework,
and the front-end is a standalone Elm project.

\subsection{Data Model}\label{subsec:datamodel}
The data model of the application is closely related to Isabelle command spans.
Source code is partitioned into blocks of text that correspond to Isabelle commands;
for each block, the semantic entities defined in that block are grouped together.
We currently differentiate between constants, facts and types for the theory entities.
When filtering for properties of theory entities in a search,
only blocks that define matching entities will be returned.

Blocks and theory entities are individually stored as Solr documents.
A Solr document consists of values for a variable set of \emph{fields}.
The list of available fields is shown in \autoref{tab:fields} --
although more fields exist for technical reasons, these are the fields relevant for querying.
Relationships between entities
(such as usage in a proof, proposition or type)
are also stored in fields, as a list of IDs.

\begin{table}[ht]
    \centering
    \begin{tabular}{l|l}
        Field & Description \\
        \hline
        Id & Unique identifier of a code block \\
        ChildId & Unique identifier of a theory entity \\
        Command & Isabelle command \\
        SourceCode & Source code, with HTML elements \\
        SourceTheory & Session-qualified theory name \\
        SourceTheoryFacet & Field for facet queries on SourceTheory  \\
        StartLine & Line at which code block starts \\
        Kind & Theory entity kind: Constant, Fact, or Type \\
        Name & Theory entity name \\
        NameFacet & Field for facet queries on Name \\
        ConstantType & Type of constants \\
        ConstantTypeFacet & Field for facet queries on ConstantType \\
        Uses & Relationship to other entities (as list of IDs)
    \end{tabular}
    \caption{Fields to be used for querying}
    \label{tab:fields}
\end{table}

\subsection{Query Specification}\label{subsec:query}
A query restricts the result set by defining \emph{filter}s on fields (available fields are listed in \autoref{tab:fields});
the result is the intersection of all filters.
Filters include standard logical connectives and nested queries.
They are defined as follows:
\begin{eqnarray*}
    \text{Filter} & := & \text{Term(String)} \\
    & | & \text{Exact(String)} \\
    & | & \text{InRange(Int, Int)} \\
    & | & \text{Not(Filter)} \\
    & | & \text{And(List Filter)} \\
    & | & \text{Or(List Filter)} \\
    & | & \text{InResult(Field, List (Field, Filter))}
\end{eqnarray*}

Filters operate on \emph{Solr terms},
which are generated from both query and indexed values by transformations that are defined in the field configurations.
For instance, values for a text field might be split on whitespace characters and transformed into lower case.

To start with, the `Term' filter matches if at least one of the Solr terms of its filter string is found in a indexed value;
additionally, a `*' wildcard is allowed.
Second, for an `Exact' filter to match, its Solr terms must be a subsequence of the indexed Solr terms.
Some sloppiness is allowed though,
i.e. as long as all Solr terms are present and reasonably close next to each other,
it is considered a sequence.
Next, the `InRange' filter only works on numerical fields,
and is inclusive for both start and end of the range.
After that, the Boolean filter connectives work as one would expect;
lastly, the 'InResults' filter executes a sub-query,
returning values of the specified field from matching elements,
and acts as a disjunction of 'Term' filters with these results.

The term splitting and transformation of the different fields is defined as part of the Solr configuration.
For the SourceText field, special characters are stored in Unicode;
a synonym mechanism allows searching by their Isabelle representations,
e.g. \texttt{\textbackslash <Longrightarrow>} or \texttt{==>} would both match $\Longrightarrow$.

Besides retrieving a list of results,
queries can also be used to retrieve \emph{facets},
i.e. a list of distinct values for a field together with their number of occurrences in the result set.
Facets are a key factor in building drill-down searches (and are extensively used in our user interface).

\subsection{Search Example}
To illustrate the search functionality, we portray a search for facts about prime numbers in the following.
This example can also be followed interactively at the `example' page on the FindFacts website.

First, the appropriate definition has to be found.
Typing in 'prime' in the main search bar on the landing page of the web UI is an obvious starting point,
but yields more than 2000 results,
which, by first glance, do not appear to be all definitions.
The facet on entity kind that has appeared allows us to select constants,
which reduces the result set to less than 150 results.
However, from the first few hits,
it is obvious that most results only use primes,
not define them.
Since we are looking for some kind of definition,
only blocks containing semantic entities with `prime' in their name are relevant --
which can be expressed by adding a filter on entity names.
A prime definition from ZF is the first hit from the remaining 60 results.
Assuming that we are looking for a HOL definition in our example,
a filter on the constant type would be useful.
Notably, the drop-down menu on the constant type filter makes it possible to scroll through the available options,
if it was not clear how the type should look like.
However, in this example, `nat => bool' is fairly obvious.
Writing down the exact type (`Nat.nat $\Rightarrow$ Hol.bool') is not necessary as the type is looked up using full-text search, not pattern unification.
A total of 18 results remain, and a facet on the Isabelle command appears with only a handful of options.
Selecting all that could be relevant (omitting locales and functions, for instance) leaves 8 results,
of which 6 are in fact definitions of primes, in different theories.
Utilizing the `used by' filter of those entities,
and then filtering for facts as well as theorem and lemma commands,
is an appropriate solution for this example.
\section{Future Work}\label{sec:future}
While we believe that the current solution is already quite useful,
there are several aspects which can be improved.

First of all, term matching by unification is not possible,
yet would be quite helpful in some situations.
This could be done in multiple ways:
One option is to build a term index, following the example of \citeauthor{mws2020kwarc};
the MathWebSearch system could also be integrated directly
(some work has already been done to import Isabelle theories in MMT, the underlying logical framework of MWS \cite{makarius2018mmt}),
but this would come with significant complexity.
An alternative and simpler option could be to first apply other filters and heuristics,
and then perform term matching directly on the reduced result set.

Syntax highlighting in the search results is not yet supported.
While this would be easy to obtain in a running PIDE session,
re-generating it from the dumped markup artifacts is not straightforward.
One could also attempt to extract the syntax-highlighting from the Isabelle HTML export.

The ordering of results can still be improved.
Currently, it only depends on the Solr-internal ranking,
which is based on how many terms match for a 'Terms' filter,
or how close the matching terms of an 'Exact' filter are.
Frequently, this rank is the same for lots of results,
for example if the search only consists of filters with single Solr terms.
The order of results then appears arbitrary.
Although the extensive search facets make it possible to further restrict to relevant results,  
a ranking that orders results based on their importance would be beneficial.
To measure importance, a graph ranking model
(e.g., Google PageRank)
could be employed on the underlying graph of theory entities.

In the short testing period prior to the writing of this paper,
integration into Isabelle has already been requested by multiple users.
Using an integrated search server, the search core can run completely standalone;
but building the whole index in the Isabelle session is way too resource-intensive.
A download mechanism for a pre-built index would be a feasible solution.
The active session could be indexed as well,
though there is no mechanism for incremental updates,
so affected theories would need to be completely re-indexed at every change.
\printbibliography{}

@online{isa-afp2020statistics,
  author={Eberl, Manuel and Klein, Gerwin and Nipkow, Tobias and Paulson, Larry and Thiemann, René},
  title={Archive of formal proofs - Statistics},
  url={https://www.isa-afp.org/statistics.html},
  urldate={2020-01-30}
}

@inproceedings{hambasan2014mathwebsearch,
  title={MathWebSearch at NTCIR-11.},
  author={Hambasan, Radu and Kohlhase, Michael and Prodescu, Corneliu-Claudiu},
  booktitle={NTCIR},
  year={2014},
  organization={Citeseer}
}

@book{ion1998mathematical,
  title={Mathematical Markup Language (MathML) 1.0 Specification},
  author={Ion, Patrick and Miner, Robert and Buswell, Stephen and Devitt, A},
  year={1998},
  publisher={World Wide Web Consortium (W3C)}
}

@online{makarius2018mmt,
  title={{Isabelle/MMT}: Export of {Isabelle} theories and import as {OMDoc} content},
  author={Wenzel, Makarius},
  url={https://sketis.net/2018/isabelle-mmt-export-of-isabelle-theories-and-import-as-omdoc-content},
  date={2018-11-09},
  urldate={2020-03-25}
}

@online{solr82,
  title={{Apache Solr} reference guide},
  organization={Apache Software Foundation},
  url={https://lucene.apache.org/solr/guide/8_2},
  date={2019-10-01},
  urldate={2020-03-27}
}

@online{lucenecore,
  title={Apache Lucene core},
  organization={Apache Software Foundation},
  year={2020},
  url={https://lucene.apache.org/core},
  urldate={2020-03-27}
}

@article{wenzel2019isabelle,
 title={The {Isabelle/Isar} Reference Manual},
  author={Wenzel, Makarius and others},
  year={2019},
  address={Technische Universität München}
}

@online{mws2020kwarc,
    title={MathWebSearch},
    organization={The KWARC Group},
    year={2020},
    author={Kohlhase, Michael and others},
    urldate={2020-03-30},
    url={https://kwarc.info/systems/mws}
}

@online{coq2019search,
    title={Vernacular commands - {Search}},
    organization={Inria and CNRS},
    year={2019},
    url={https://coq.inria.fr/refman/proof-engine/vernacular-commands.html#coq:cmd.search},
    urldate={2020-03-31}
}

\end{document}